\documentclass[twocolumn,showpacs,preprintnumbers,amsmath,amssymb]{revtex4}

\usepackage{graphicx}
\usepackage{dcolumn}
\usepackage{bm}
\usepackage{epsfig}
\usepackage[mathscr]{euscript}
\usepackage{color}

\newenvironment{Notes}
{\begin{quote}\small\tt Note to Ioan: \ }
{\end{quote}}

\newcommand{\beq}{\begin{equation}}
\newcommand{\eeq}{\end{equation}}
\newcommand{\bno}{\begin{Notes}}
\newcommand{\eno}{\end{Notes}\noindent}

\begin{document}

\title{Advancements in Milestoning I: Accelerated Milestoning via ``Wind'' Assisted Re-weighted Milestoning (WARM)}

\author{Gianmarc Grazioli}
\author{Ioan Andricioaei}
 \email{andricio@uci.edu}
\affiliation{Department of Chemistry, University of California, Irvine, CA 92697}

\date{\today}

\begin{abstract}
\noindent
The Milestoning algorithm created by Ron Elber et al. is a method for determining the time scale of processes too complex to be studied using brute force simulation methods. The fundamental objects implemented in the Milestoning algorithm are the first passage time distributions $K_{AB}(\tau)$ between adjacent protein configuration milestones $A$ and $B$. The method proposed herein aims to further enhance Milestoning by employing an artificial applied force, akin to wind, which pushes the trajectories from their initial states to their final states, and subsequently re-weights the trajectories to yield the true first passage time distributions $K_{AB}(\tau)$ in a fraction of the computation time required for unassisted calculations. The re-weighting method, rooted in Ito's stochastic calculus,  was adapted from previous work by Andricioaei et al. The theoretical basis for this technique and numerical examples are presented. 
 
\end{abstract}

\pacs{Valid PACS appear here}
\keywords{molecular dynamics}

\maketitle

\section{Introduction}
The task of calculating kinetic properties from molecular dynamics simulations is a complex problem of considerable interest \cite{dellago1998transition} \cite{dellago1999calculation}. In contrast to computational methods designed for equilibrium calculations, in which the basic observables are thermodynamic averages over conformational points (structures) generated over an invariant measure without the need to obey exact dynamical equations, studies of kinetics require physically correct time-ordered trajectories to obtain time-correlation functions as the basic objects \cite{Elber2011}. Since each time-correlation function describes the relaxation under investigation as an average over all relevant trajectories, adequate sampling for accurate calculation of long-time kinetics can quickly become computationally intractable via direct simulation \cite{Berendsen1998}. This is because a direct, brute force method of this type would require sufficiently long simulation times such that the system would be able to transition between states of interest enough times that a statistically significant distribution of first passage times could be generated. Several computational methods have been developed to address the challenge of calculating chemical kinetics, starting with the venerable transition state theory (TST) \cite{Eyring} \cite{Wigner1938}, and continuing with more recent developments, such as, transition path sampling (TPS) \cite{Bolhuis2002}, transition path theory (TPT) \cite{VandenEijnden2009}, and transition interface sampling (TiS) \cite{TiS}. Although transition state theory has been successfully used in the determination of the kinetics for many systems with well-defined reactant and product states, for which the ``dynamical bottleneck" can be identified \cite{truhlar1996current}, there are many interesting problems in biophysics, and elsewhere, for which these assumptions do not hold. In contrast, transition path sampling approaches require no intuition for reaction mechanisms or advance knowledge of transition state, although the requirement of a "dynamical bottleneck" does persist \cite{Bolhuis2002} \cite{vanden2010transition}. In the same category of methods is the milestoning algorithm created by Ron Elber et al., which is a method for calculating kinetic properties, where the fundamental objects are the first passage time distributions $K_{AB}(\tau)$ between adjacent protein configuration milestone states (configurations $A$ and $B$ in this case), where the milestone states do not necessarily need to be meta-stable states as in transition state theory. The key feature of the milestoning method is that long trajectory pathways for large scale configuration changes can be broken up into shorter trajectories for which a linear network of transition probabilities between milestones can be devised. The aforementioned linear networks of transition probabilities can then be solved for such quantities as first passage time between any pair of milestones, including those at the extreme ends of the space, and the flux through a given milestone, $s$, as a function of time, written as $P_s(t)$ (equation \ref{m}). Some of the key gains from this treatment are that breaking up these long trajectory pathways into a network of shorter trajectories leads to increased sampling of the would-be under-sampled areas, and that gains in computational efficiency are possible due to the capacity to run these short trajectories in parallel \cite{Elber2004}. In practice, previous milestoning calculations have been limited to calculating the constant flux values representative of the system at equilibrium, which can be thought of as the long time flux values $\lim_{t \to \infty}P_s(t)$. A method for calculating the time-dependent flux through a given milestone $P_s(t)$ can be found in the companion paper to this article, also in this publication \cite{Grazioli2}. The aim of the technique we present in this paper is to increase the computational speed of the milestoning method via the addition of an artificial constant force ($\mathscr{F}_{wind}$) along the vector pointing from the initial state to the final state for each pair of milestones in the simulation, causing the system to arrive at the destination configuration in far fewer time steps than if it were left to Brownian dynamics alone. The key idea which makes this possible is the use of a re-weighting function we have introduced previously \cite{Andricioaei2007} \cite{Xing} \cite{Daun} \cite{Yassin}, which generates a re-weighting coefficient for each trajectory, thus allowing the true distribution of first passage times to be recovered from the artificially accelerated trajectories. Preliminary calculations conducted on model systems, described in the Numerical Demonstration section, have demonstrated a computation time speedup by a factor of nearly $40$ using this method.  

\section{Theory}
\noindent 

The quantity of most fundamental importance in milestoning is the flux through a given milestone, for which the equation is \cite{Elber2004}:

\begin{equation}
\label{m}
P_s(t) = \int_0^t Q_s(t')\left[ 1-\int_0^{t-t'}K_s(\tau)d\tau \right]dt' , \nonumber
\end{equation}

\begin{equation}
Q_s(t) = 2 \delta(t)P_s(0) + \int_0^t Q_{s\pm1}(t'')K^{\mp}_{s\pm1}(t-t'')dt''
\end{equation}
\noindent
where $P_{s}(t)$ is the probability of being at milestone $s$ at time $t$, (or, more specifically, arriving at time $t'$ and not leaving before time $t$ \cite{Elber2004}), and $Q_{s}(t)$ is the probability of a transition to milestone $s$ at time t.  $K_s(\tau)$ indicates the probability of transitioning out of milestone $s$ given an incubation time of $\tau$, thus $\int_0^{t-t'}K_s(\tau)d\tau$ is the probability of an exit from milestone $s$ anytime between $0$ and $t-t'$, which makes $1-\int_0^{t-t'}K_s(\tau)d\tau$ the probability of there \emph{not} being an exit from milestone $s$ over that same time period. Since the probability of two independent simultaneous events happening concurrently is the product of the two events, the equation for $P_s(t)$ is simply integrating the concurrent probabilities of arriving at milestone $s$ and not leaving over the time frame from time $0$ to $t$. Dissecting the meaning of $Q_{s}(t)$ (14), the first term, $2 \delta(t)P_s(0)$, simply represents the probability that the system is already occupying milestone $s$ at time $t = 0$, where the factor of 2 is present since the $\delta$ function is centered at zero, meaning only half of its area would be counted without this factor. $Q_{s\pm1}(t'')$ is the probability that the system transitioned into one of the two milestones adjacent to $s$ at an earlier time $t''$. $K^{\mp}_{s\pm1}(t-t'')$ is the probability of a transition from milestones $s\pm 1$ into milestone $s$. Thus the second term of the second line of equation 14 is another concurrent probability: the probability of the system entering an adjacent milestone at an earlier time, and then transitioning into milestone $s$ between time $t$ and $0$. It is important to note that all functions $P_s(t)$ and $Q_s(t)$ are calculated using the respective values of $K_s(\tau)$ between adjacent milestones, thus the set of $K_s(\tau)$ between all milestones of interest contains all the information needed to calculate kinetics using the milestoning method. 

The important connection to make in regard to combining the milestoning method with re-weighting of artificially accelerated trajectories is that a $K$ function between two milestones $x = A$ and $x = B$, $K_{AB}(\tau)$, is nothing more than a probability distribution as a function of lifetime describing the conditional probability that a system found in state $A$ at time $t = 0$ will be found, for the first time, in state $B$ at time $t = \tau$:

\begin{equation}
K_{AB}(\tau) = P(x_B, \tau | x_A, 0)
\end{equation}    


\noindent Given this relationship, we can now begin to make the connection between milestoning and re-weighting of artificially accelerated trajectories. Assuming Langevin dynamics with the addition of a $wind$ force:

\begin{equation}
m\ddot{x} = -\gamma m \dot{x} -\nabla V(x) + \xi (t) + \mathscr{F}_{wind}
\end{equation}

\noindent where $\gamma$ is the friction coefficient, $V(x)$ is the potential, $\xi (t)$ is the random force, and $\mathscr{F}_{wind}$ is a constant force applied in the direction of the goal milestone for each run; conditional probabilities reflecting first passage transitions from milestone $A$ to $B$ can be expressed as:

\begin{equation}
P(x_B, \tau | x_A, 0) = \int D \xi W[\xi] \delta (x(\tau) - x_B)
\end{equation}
\noindent
In this equation, $W[\xi(t)]$ is the probability distribution representing the joint probabilities of all possible series of random kicks, so multiplying by the delta function $\delta (x(\tau) - x_B)$ and integrating selects for only the portion of the distribution which represents a series of random kicks which results in a transition from state $A$ to state $B$ given an incubation time $\tau$. It then follows suit that the integral in this equation is simply the expectation value for the probability of a transition from $A$ to $B$ for each incubation time point $\tau$, which, again, is equivalent to $K_{AB}(\tau)$. Because of fluctuation dissipation, $\langle \xi(t) \xi(t') \rangle = 2 k_B T m \gamma$, the random force in Langevin dynamics is a Gaussian distribution with variance $w \equiv 2 k_B T m \gamma$. Thus it can be show that:

\begin{equation}
W[\xi(t)] = \exp (-\frac{1}{2w}\int_0^t \xi(t')^2 dt')
\end{equation}

With $W[\xi(t)]$, our weighting function for joint probabilities of random kick sequences in terms of our random force $\xi(t)$ in hand, we can now write the noise history $\xi(t)$ in terms of the trajectory $x(t)$ it generates:

\begin{equation}
\xi(t)^2 = (m\ddot{x} + \gamma m \dot{x} + \nabla V(x) - \mathscr{F}_{wind})^2
\end{equation}
\noindent
We are ultimately interested in measuring conditional probability distributions in configuration space, $x(t)$, not random force space, $\xi(t)$, but since the Jacobian is built into the measure, $x$, we can define $S[x(t)]$ thusly:


\begin{equation}
S[x(t)] \equiv \xi(t)^2 = (m\ddot{x} + \gamma m \dot{x} + \nabla V(x) - \mathscr{F}_{wind})^2
\end{equation}

\noindent
Then, using the Ito formalism for stochastic calculus, we can express our conditional probability using the Wiener formalism of path integrals \cite{Kleinert2004} as:

\begin{equation}
P(x_B, \tau | x_A, 0) = \int_{(x_A, 0)}^{(x_B, \tau)} D x \exp \left( -\frac{S[x(t)]}{2w} \right)
\end{equation}
\noindent

In this form, it is clear that the exponential function represents the weighting function for the trajectory $x(t)$:

\begin{equation}
W[x(t)] \equiv \exp \left( -\frac{S[x(t)]}{2w} \right)
\end{equation}
\noindent
With the weight of each trajectory, $W[x(t)]$, now formally defined, we can define the re-weighting factor for obtaining the true weight of an artificially accelerated trajectory as:

\begin{equation}
\frac{W[x(t)]}{W_f[x(t)]} = \exp \left( -\frac{S[x(t)] - S_f[x(t)]}{2w}\right)
\end{equation}
\noindent 
where the $f$ subscript indicates a function generated under the influence of the artificial $\mathscr{F}_{wind}$ force. In practice, once a trajectory $x(t)$ is generated (in the presence of the wind force), the actions are calculated in discrete numerical form using:

\begin{equation}
S_f[x(t)] \approx \sum_i \left(  m\frac{\Delta \nu_i}{\Delta t} + m\gamma \frac{\Delta x_i}{\Delta t} + \nabla V_i - \mathscr{F}_{wind}  \right)^2 \Delta t \nonumber
\end{equation}

\begin{equation}
S[x(t)] \approx \sum_i \left(  m\frac{\Delta \nu_i}{\Delta t} + m\gamma \frac{\Delta x_i}{\Delta t} + \nabla V_i   \right)^2 \Delta t
\label{x}
\end{equation}
\noindent

The re-weighting factor is calculated from Eq. (10) and stored in an array. When post-processing to compute the $K_{AB}(\tau)$ distribution for a particular pair of milestones $A$ and $B$ by histogramming trajectories by lifetime $\tau$, instead of adding 1 to a particular bin each time the lifetime of a particular trajectory falls within the bounds of that bin, the weight $W[x(t)]$ corresponding to that trajectory is instead added. It is clear from equation 10 that as $S_f[x(t)]$ for the artificially accelerated trajectory approaches $S[x(t)]$, the weight of the trajectory approaches unity, thus the method reduces to an unweighted histogram for $\mathscr{F}_{wind} = 0$ as it should. 


\section{Numerical Demonstration}

\subsection*{Model System in One Dimension}
A simple two well potential (see inset of figure 7) of equation $y = (x-1)^2(x+1)^2$ was chosen to be the model system upon which the wind-assisted milestoning methodology could be developed. In running wind-assisted milestoning, the potential to which the particle is being subjected is first divided into any number of milestones, in this case, 7 milestones, thus 6 separate spaces. Next, numerous Langevin trajectories are run both from left to right and right to left between each pair of adjacent milestones. The number of time steps required to go from the starting milestone to the destination milestone for each trial of each pair and the weight of each trajectory is then stored in an array as mentioned in the theory section. As shown in figures 1 and 2, this method has brought about a more than tenfold speedup in computation time with very little sacrifice in terms of accuracy. 

\begin{figure}[h]
  \centerline{\epsfig{figure=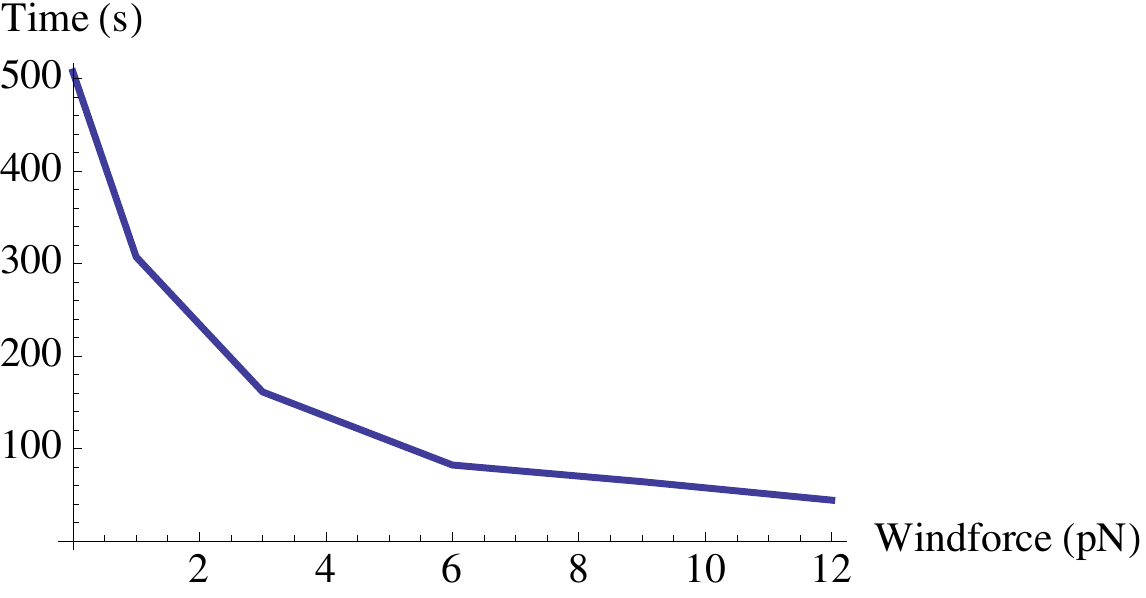,width=3.5in}}
  \caption{Shown here is calculation time as a function of the $\mathscr{F}_{wind}$ force for all $K(\tau)$ distributions in both directions for six subspaces ranging from -2 to 2 on the bistable harmonic potential. The calculation took 507 seconds for $\mathscr{F}_{wind} = 0 pN$ and just 44 seconds for $\mathscr{F}_{wind} = 12 pN$}.
  \label{fig:US}
\end{figure}

\begin{figure}[h]
  \centerline{\epsfig{figure=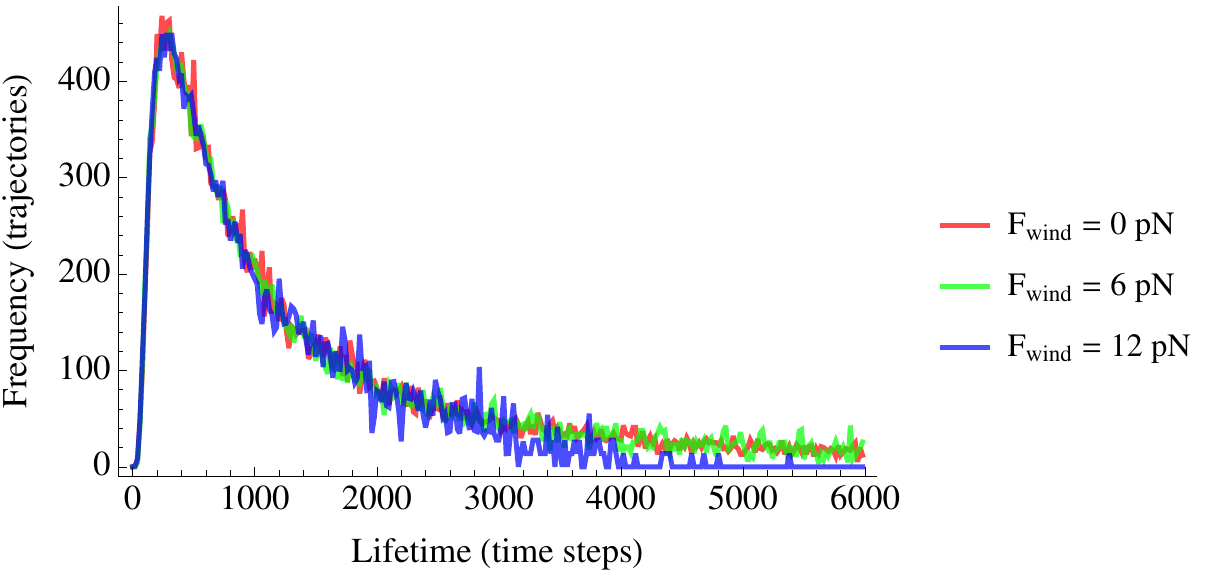,width=3.65in}} 
  \caption{Shown in this figure is the transition probability distribution $K_{23}(\tau)$, i.e. the transition probability from milestone 2 to milestone 3 as a function of lifetime, calculated using $\mathscr{F}_{wind}$ forces ranging from 0 to 12 pN. The plots indicate that the rapid decrease in computation time due to the added $\mathscr{F}_{wind}$ force has almost no effect on accuracy.}
  \label{fig:F} 
\end{figure}

\subsection*{Model System in One Dimension with Distortion}

\begin{figure}[h]
  \centerline{\epsfig{figure=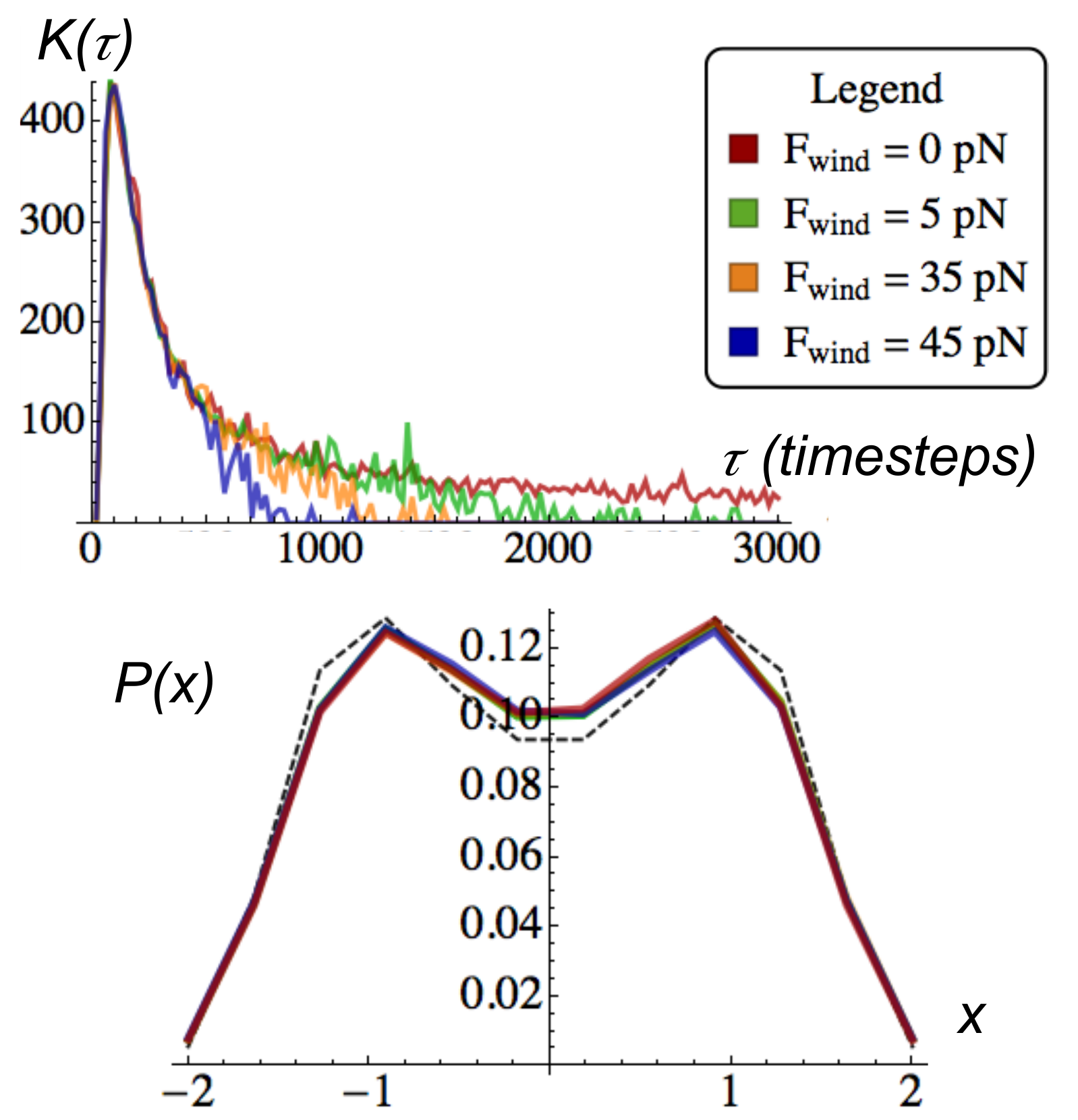,width=3.3in}} 
  \caption{The plot at the top of this figure shows plots for one of the transition probability distributions $K(\tau)$ for the bistable 1D potential with different values of $\mathscr{F}_{wind}$ implemented. Note that although the distributions distort considerably for higher values of $\tau$ when the system is pushed with high magnitude $\mathscr{F}_{wind}$, the equilibrium flux values in the plot below remain fairly constant. The color scheme legend applies to both plots.}
  \label{P} 
\end{figure}

\begin{figure}[h]
  \centerline{\epsfig{figure=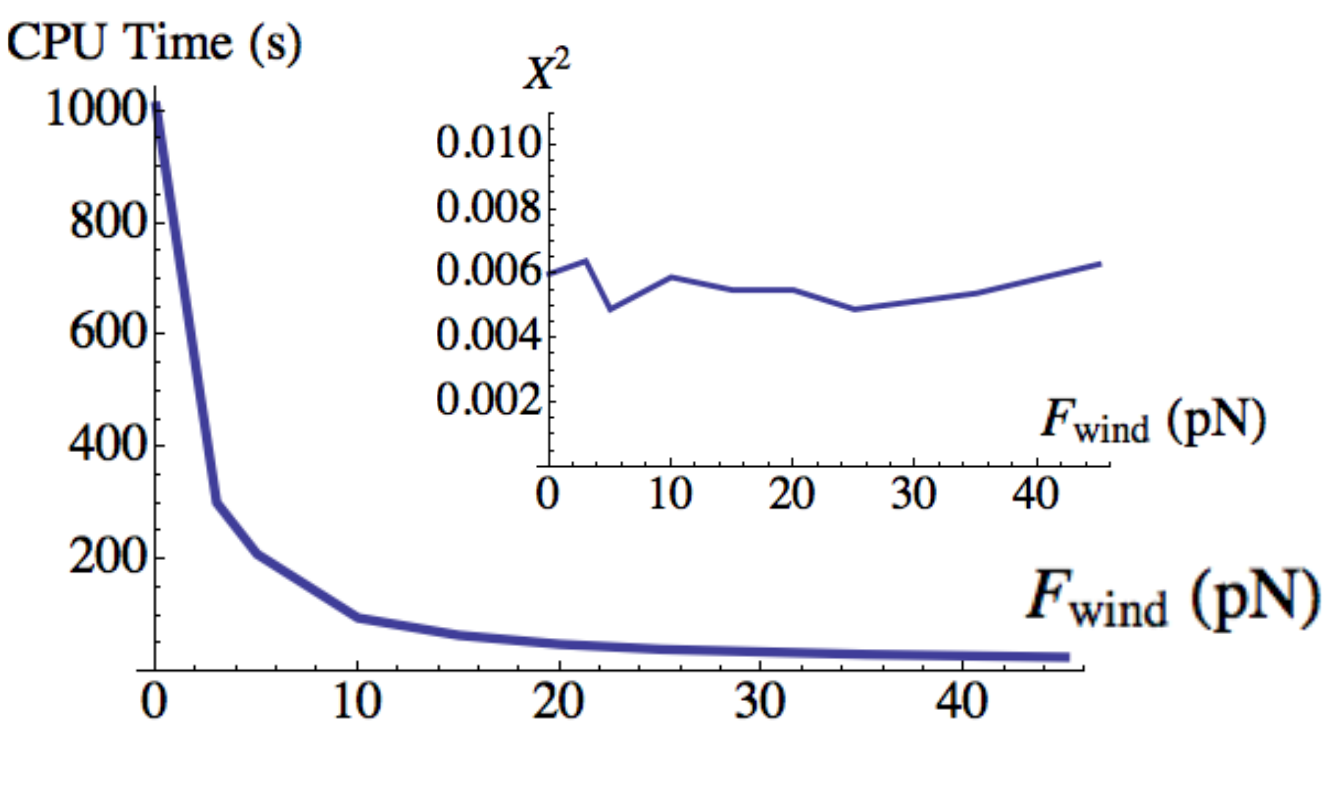,width=3.5in}} 
  \caption{Here we show effects of applying higher magnitude $\mathscr{F}_{wind}$ which are strong enough to significantly distort the $K(\tau)$ functions. This figure facilitates a direct comparison of gain in computational speed with the accuracy of the equilibrium flux values (measured as $X^2$). Note that while there is no appreciable change in accuracy, calculation time drops from 1109 s to 26 s, a speedup by a factor of nearly 40.}
  \label{Q} 
\end{figure}

Thus far, we have approached the WARM method from the standpoint of speeding up the calculation by pushing $\mathscr{F}_{wind}$ until the $K(\tau)$ functions begin to distort. Here we will explore the possibility that even slightly distorted $K(\tau)$ functions can yield useful information, allowing for even greater computational speedup. The flux value for a given milestone $s$, $P_s(t)$, should approach the probability predicted by the Boltzmann distribution generated from configurational partition function as time approaches infinity. Given a discrete space in $x$, subject to our 1D potential $y = (x-1)^2(x+1)^2$, the Boltzmann distribution function can be obtained in the usual way, shown in equation \ref{B}, below:

\begin{equation}
\lim_{t \to \infty} P_s(t) = \frac{e^{-\beta U(x_s)}}{ \sum_{n=1}^{N_S} e^{-\beta U(x_n)}}
\label{B}
\end{equation}
\noindent

where $N_s$ is the total number of milestone configurations, and $x_n$ signifies the spatial position of each milestone. This discrete space approximation for the equilibrium flux values is utilized below as a test for accuracy in figure \ref{P} (dashed lines). The numerical demonstration in this section consists of dividing the space for the bistable 1D potential between $x = -2$ and $x = 2$ into 11 subspaces bounded by 12 milestones. First hitting trajectories were run between each pair of adjacent milestones, and then each pair of $K(\tau)$ functions describing a transition away from each milestone were normalized (e.g. for milestone 3, all trajectories must terminate at either milestone 2 or milestone 4, therefore $\int_0^\infty K_{32}(\tau) d\tau + \int_0^\infty K_{34}(\tau) d\tau = 1$). The normalized $K(\tau)$ functions are then integrated over all time, and these values are placed in a matrix, \textbf{K}, of equilibrium transition probabilities. The equilibrium flux values for the vector representing the set of milestones, \textbf{P}, is then found by numerically solving for the eigenvector: \textbf{P$\cdot$K} = \textbf{P} \cite{ElberWeb}. By using this method to determine equilibrium flux values, it is demonstrated in figures \ref{P} and \ref{Q}, that even when $\mathscr{F}_{wind}$ is set to a value strong enough to distort the $K(\tau)$ functions, accurate equilibrium flux values can still be calculated. 

\subsection*{Model Systems in Two Dimensions}
Two additional test systems for the WARM technique were implemented for further validating the method in two dimensions. Both systems have double well shapes, however for one well, the barrier to transition from one well to the other is primarily energetic, while the other is primarily entropic (see figure 3 below). The potential with the energetic barrier is a generalization of the 1D potential from the previous section to two dimensions, and the potential with the entropic barrier is the same potential implemented by Elber and Faradjian in their original paper on milestoning \cite{Elber2004}. As can be seen in the data below, the WARM method successfully re-weighted first passage time distributions ($K(\tau)$) generated using artificially accelerated trajectories to yield the true first passage time distributions which would have resulted from trajectories in the absence of the $wind$ force. In both cases, the method achieved more than 60\% faster computation times with very little sacrifice in terms of accuracy.

\begin{figure}[h]
  \centerline{\epsfig{figure=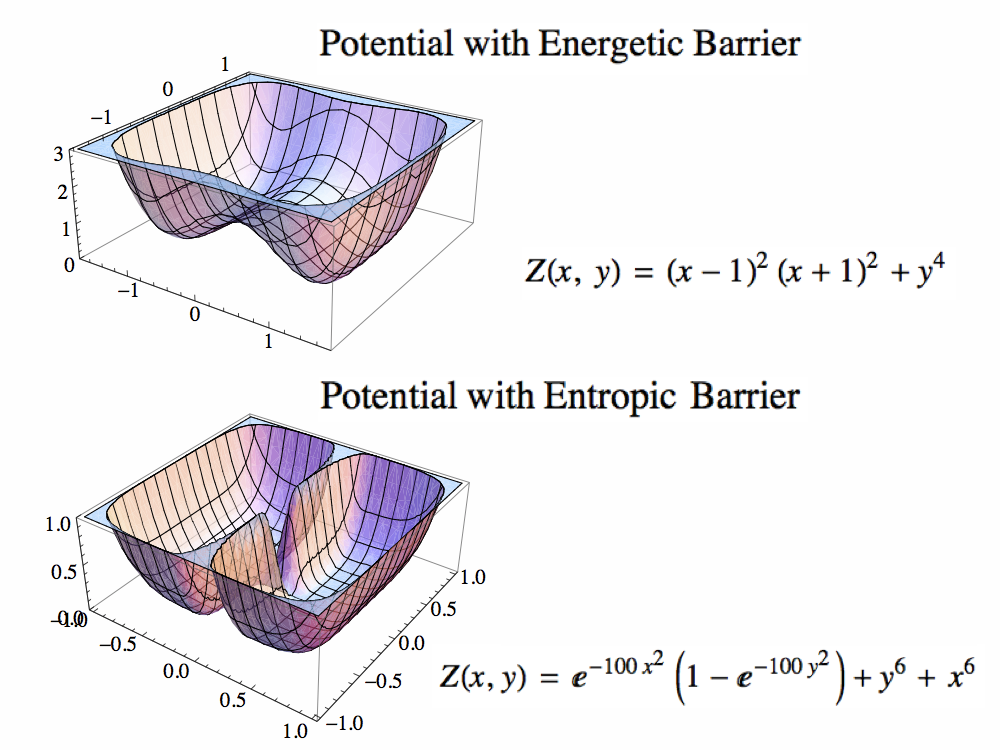,width=3.5in}} 
  \caption{Show here are the potentials used in the 2D WARM calculations. In the first case, the primary barrier to crossing from one well to the other is the height of the barrier relative to the strength of the ``kicks" from the random force in the Langevin equation. In the second potential \cite{Elber2004}, the barrier to crossing between wells is entropic, in that a trajectory which results in a transition between wells must find its way through the gap at the center, i.e. the likelihood of a transition is not limited by any sort of uphill battle, but instead by decreased degeneracy in the number of possible trajectories which result in a transition. }
  \label{fig:F} 
\end{figure}

\begin{figure}[h]
  \centerline{\epsfig{figure=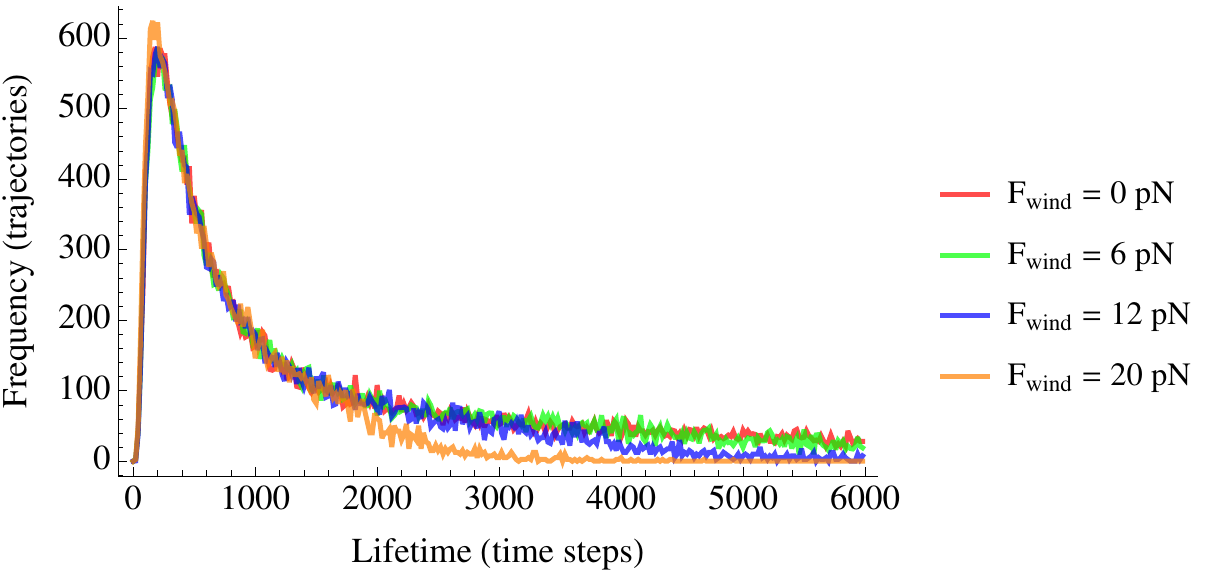,width=3.65in}} 
  \caption{Shown in this figure is the transition probability distribution $K_{12}(\tau)$, i.e. the transition probability from milestone 1 (the line $x = -1$) to milestone 2 (the line $x = 0$) on the the 2D potential with the energetic barrier as a function of lifetime, calculated using $\mathscr{F}_{wind}$ forces ranging from 0 to 12 pN. The plots indicate that the rapid decrease in computation time due to the added $wind$ force has almost no effect on accuracy. }
  \label{fig:F} 
\end{figure}

\begin{figure}[h]
  \centerline{\epsfig{figure=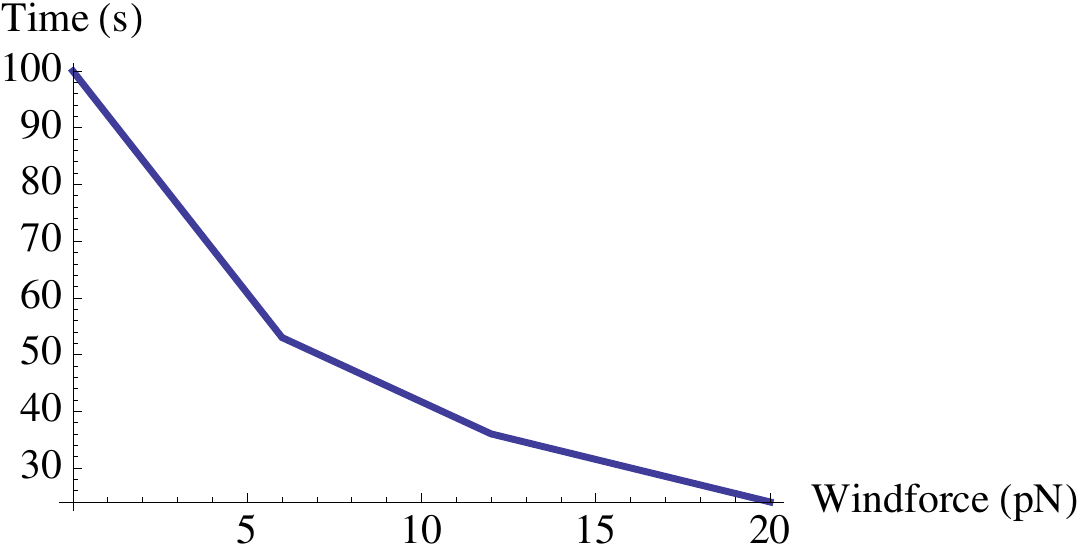,width=3.5in}}
  \caption{Shown here is calculation time as a function of the $\mathscr{F}_{wind}$ force for all $K(\tau)$ distributions in both directions for two subspaces ranging from -1 to 1 on the x axis of the 2D potential with the energetic barrier. All trajectories were run using $\beta = .123$. The highest value of $\mathscr{F}_{wind}$ yielded a faster computation time by a factor of  4.17 than the unassisted calculation with very little distortion to the $K(\tau)$ function.}
  \label{fig:US}
\end{figure}

\begin{figure}[h]
  \centerline{\epsfig{figure=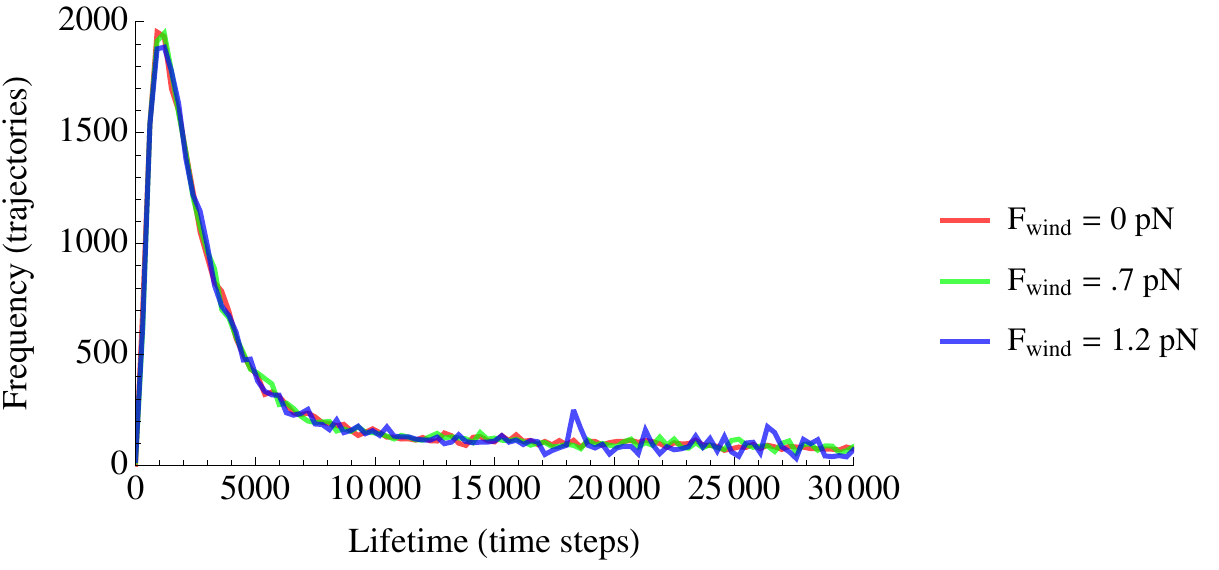,width=3.65in}} 
  \caption{Shown in this figure is the transition probability distribution $K_{12}(\tau)$, i.e. the transition probability from milestone 1 (the line $x = -.5$) to milestone 2 (the line $x = 0$) on the the 2D potential with the entropic barrier as a function of lifetime, calculated using $\mathscr{F}_{wind}$ forces ranging from 0 to 1 pN. The plots indicate that the rapid decrease in computation time due to the added $\mathscr{F}_{wind}$ force has almost no effect on accuracy.}
  \label{fig:F} 
\end{figure}

\begin{figure}[h]
  \centerline{\epsfig{figure=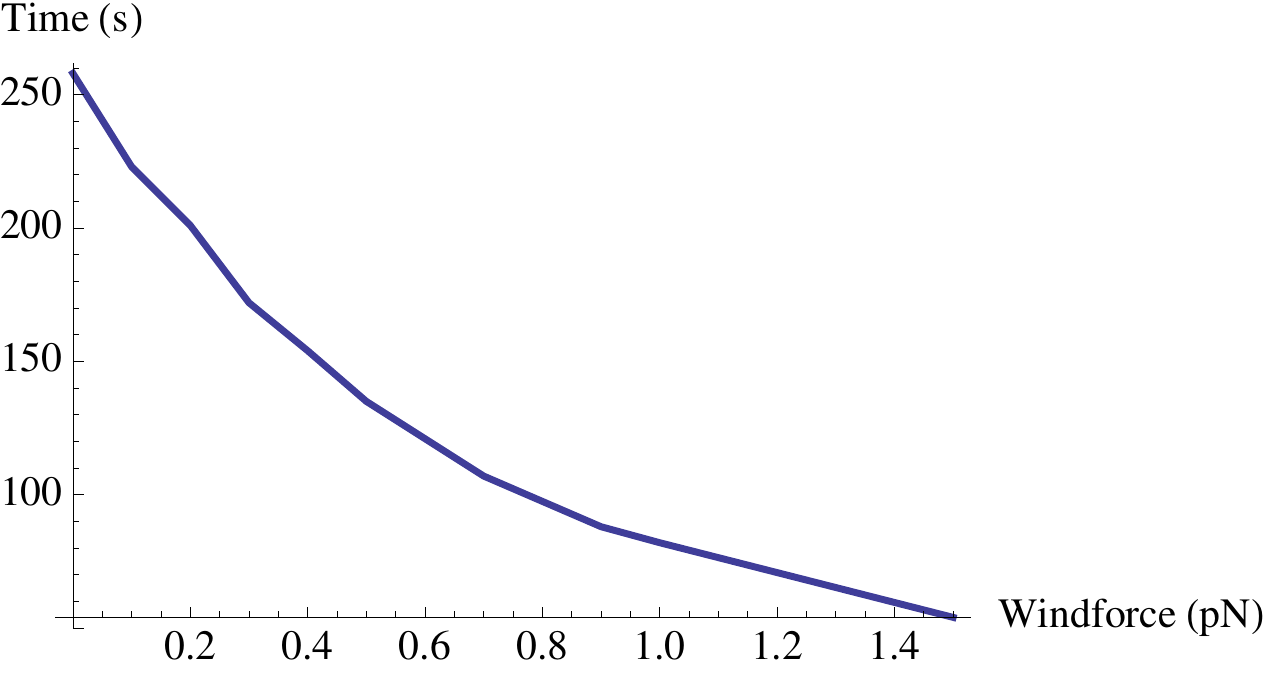,width=3.5in}}
  \caption{Shown here is calculation time as a function of the $\mathscr{F}_{wind}$ force for all $K(\tau)$ distributions in both directions for two subspaces ranging from -.5 to .5 on the x axis of the 2D potential with the entropic barrier. All trajectories were run using $\beta = 3.0$ so as to ensure that transitions over the barrier instead of through the small gap were highly unlikely. The highest value of $\mathscr{F}_{wind}$ yielded a faster computation time by a factor of 4.78 than the unassisted calculation with almost no distortion to the $K(\tau)$ function.}.
  \label{fig:US}
\end{figure}

\subsection*{Model System in Eleven Dimensions}

\begin{figure}[h]
  \centerline{\epsfig{figure=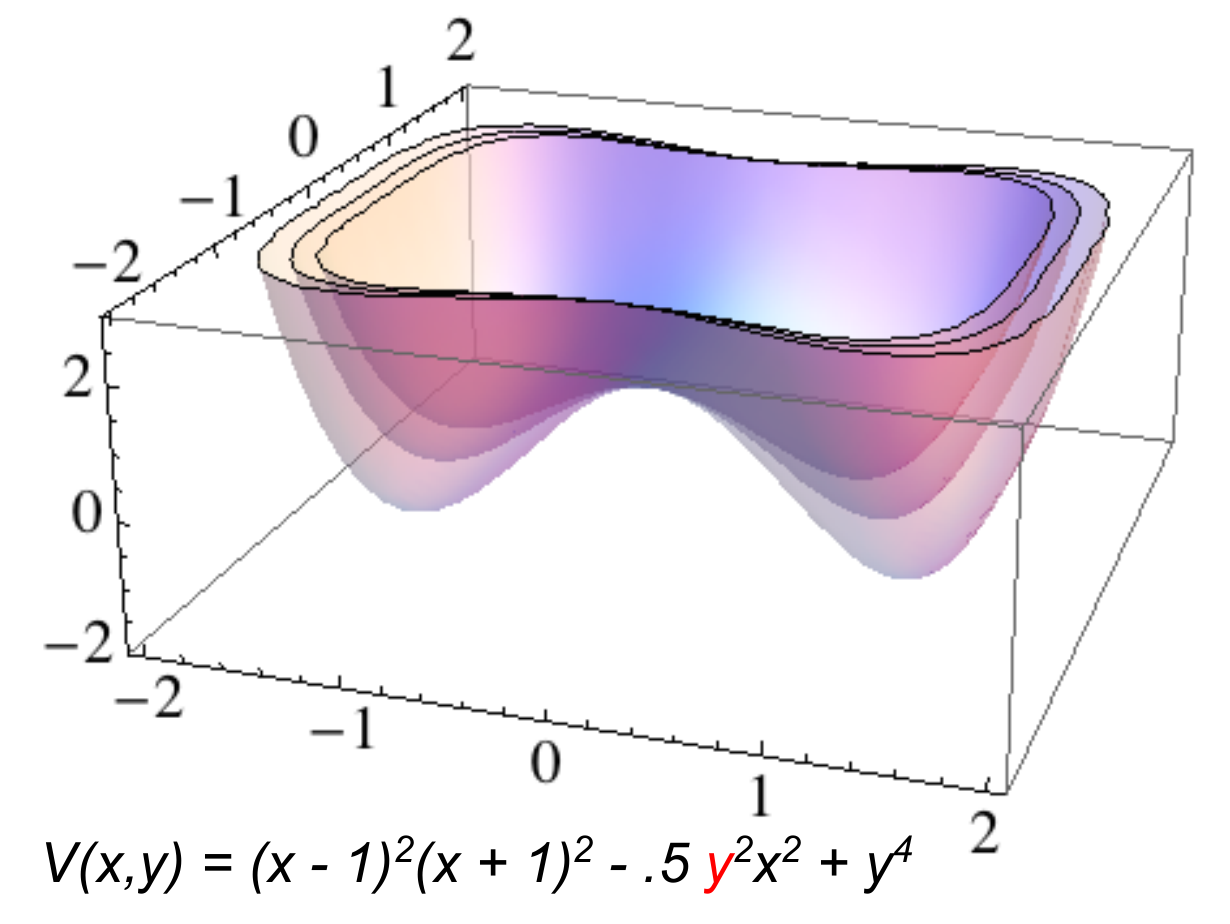,width=3.5in}} 
  \caption{Show here is a 2D representation of the 11D coupled potential. The $y$ in the second term (red) has been left as a parameter in this plot. The surfaces shown are for values for the parametric $y$ of $0, \pm 1,$ and $\pm 1.5$, where the deepest well corresponds to $y  = 1.5$ and the shallowest corresponds to parametric $y = 0$. Just as the well becomes deeper, the further from the system wanders from the origin in the $y$ direction in this 2D model, the 11D system also encounters deeper wells in the $x_n$ dimensions the further it wanders from the origin in each $x_n$ dimension.}
  \label{cp} 
\end{figure}

In order to demonstrate that the WARM method possesses no inherent limitations due to scaling, the method was applied to an $11$ dimensional hyperspace. For this model, the $11D$ potential was defined as:

\begin{equation}
V(x_1, x_2,..., x_{11}) = (x_1 - 1)^2(x_1+1)^2 -  \frac{1}{2}\sum_{n = 2}^{11} x_n^2 x_1^2 + \sum_{n = 2}^{11} x_n^4  
\end{equation}

\noindent where the first term is the same bistable potential in $x_1$ used in the first one dimensional example, the second term couples motion in the $10$ dimensions orthogonal to barrier height in $x_1$, and the third term simply confines the system to a reasonably sized configurational space using a quartic potential. In order to develop some intuition for this potential, see figure \ref{cp}, then just imagine that there are nine other dimensions which have the same effect as $y$ on barrier height in $x_1$. 

Accordingly, the milestones must be defined as hyperplanes, given the general definition of a hyperplane:

\begin{equation}
a_1 x_1 + a_2 x_2 + a_3 x_3 ... a_n x_n = b
\end{equation}
\noindent

\begin{figure}[h]
  \centerline{\epsfig{figure=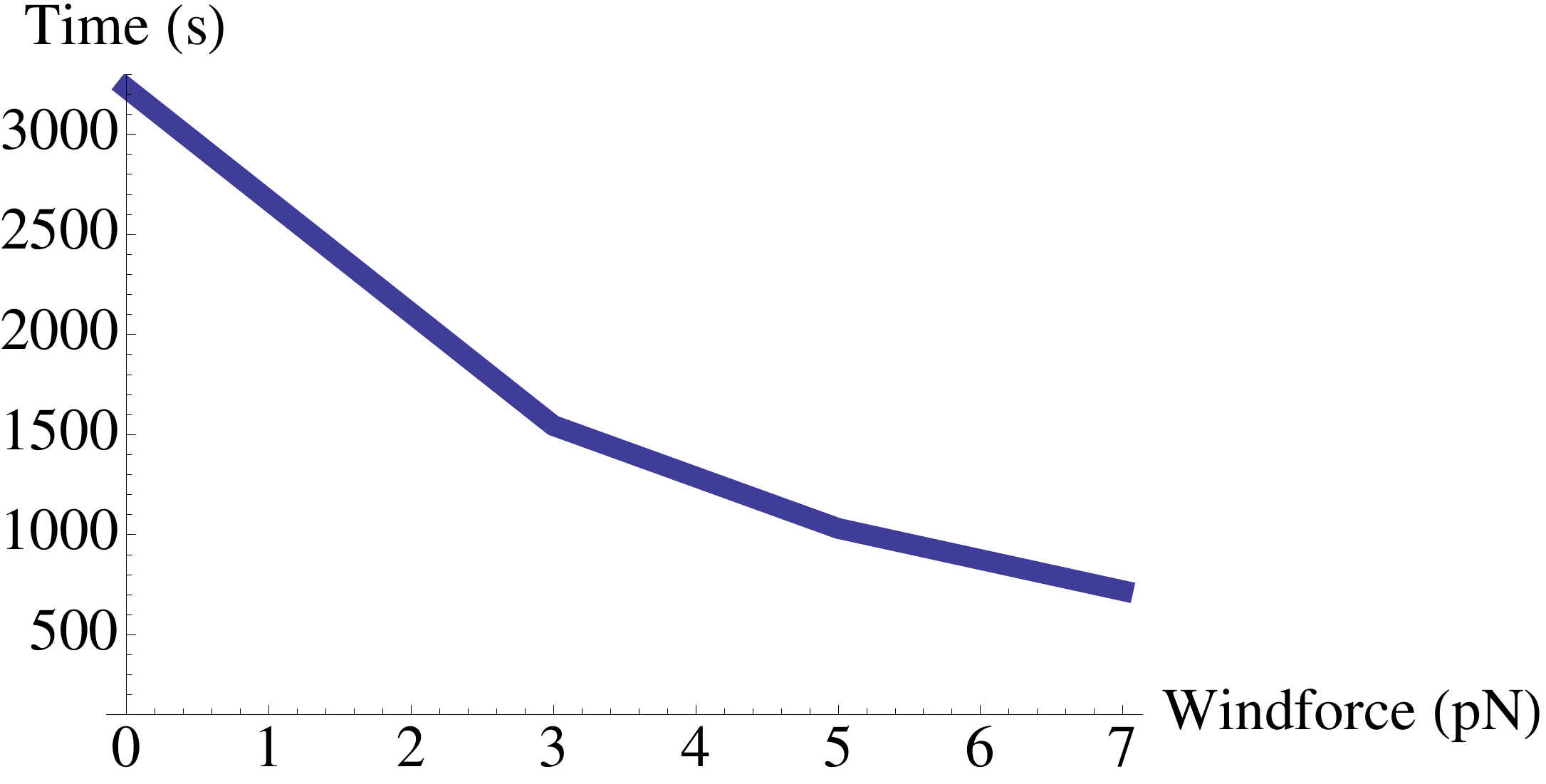,width=3.5in}}
  \caption{This plot shows CPU time as a function of the magnitude of the $\mathscr{F}_{wind}$ in 11D. The maximum speedup measured was a factor of 4.5.}
  \label{R}
\end{figure}

\begin{figure}[h]
  \centerline{\epsfig{figure=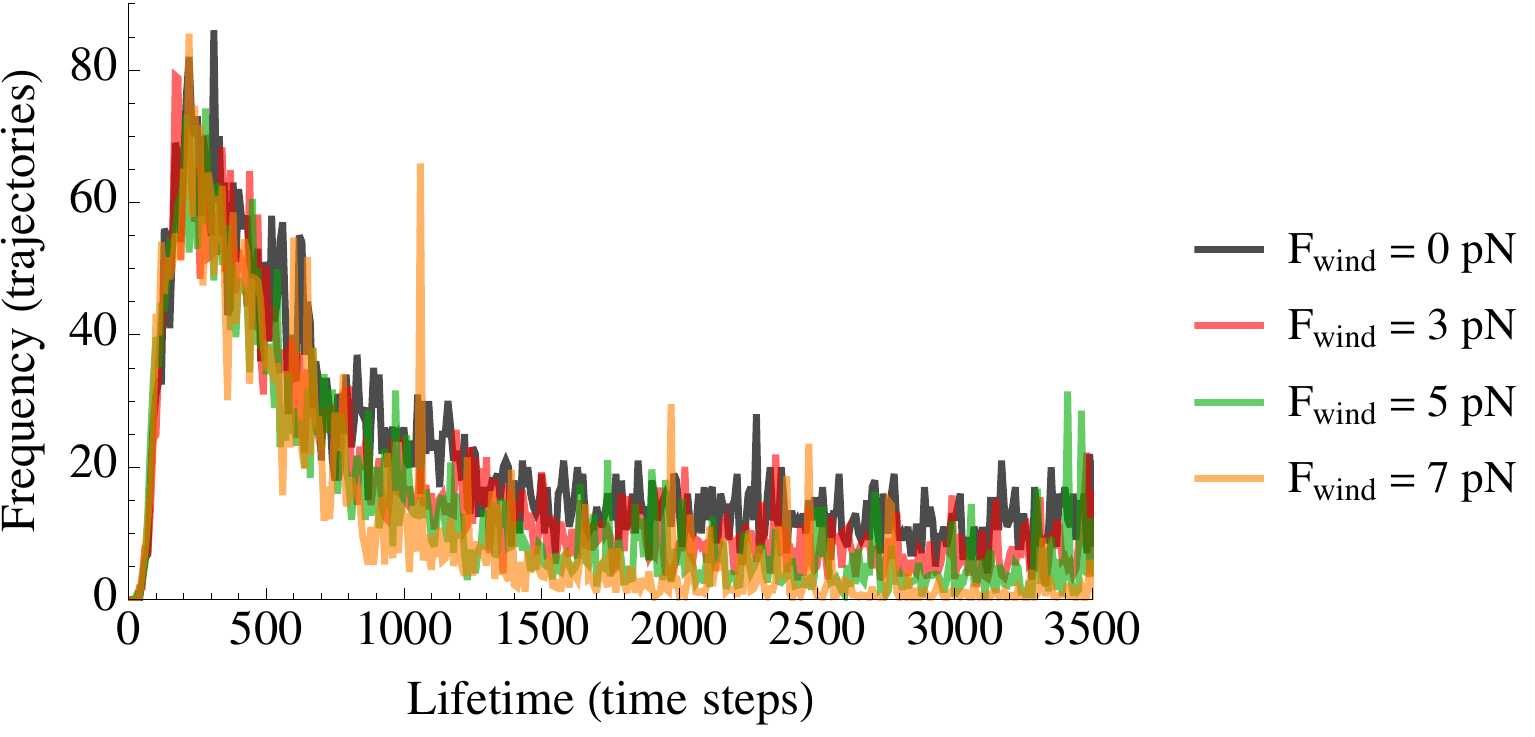,width=3.7in}} 
  \caption{Shown in this figure are the $K(\tau)$ functions generated for each data point in the CPU time vs. $\mathscr{F}_{wind}$ plot for the 11D system.}
  \label{T} 
\end{figure}

To keep things simple, we set $a_2$ through $a_{11}$ equal to zero, and $a_1 = 1$, allowing us to define two hyperplanes as  $x_1 = -1$ and $x_1 = 1$. In this scenario, the features of interest are the transitions between the wells at $x_1 = -1$ and $x_1 = 1$, thus the 11D $\mathscr{F}_{wind}$ is applied with zero components in all dimensions except for $x_1$ where it is used to push the system over the barrier between wells. The WARM method was successfully applied to this 11 dimensional potential, and a speedup by a factor of 4.5 was observed (figures \ref{R} and \ref{T}).   

\subsection*{Wind Force as a Vector Field}
\begin{figure}[h]
  \centerline{\epsfig{figure=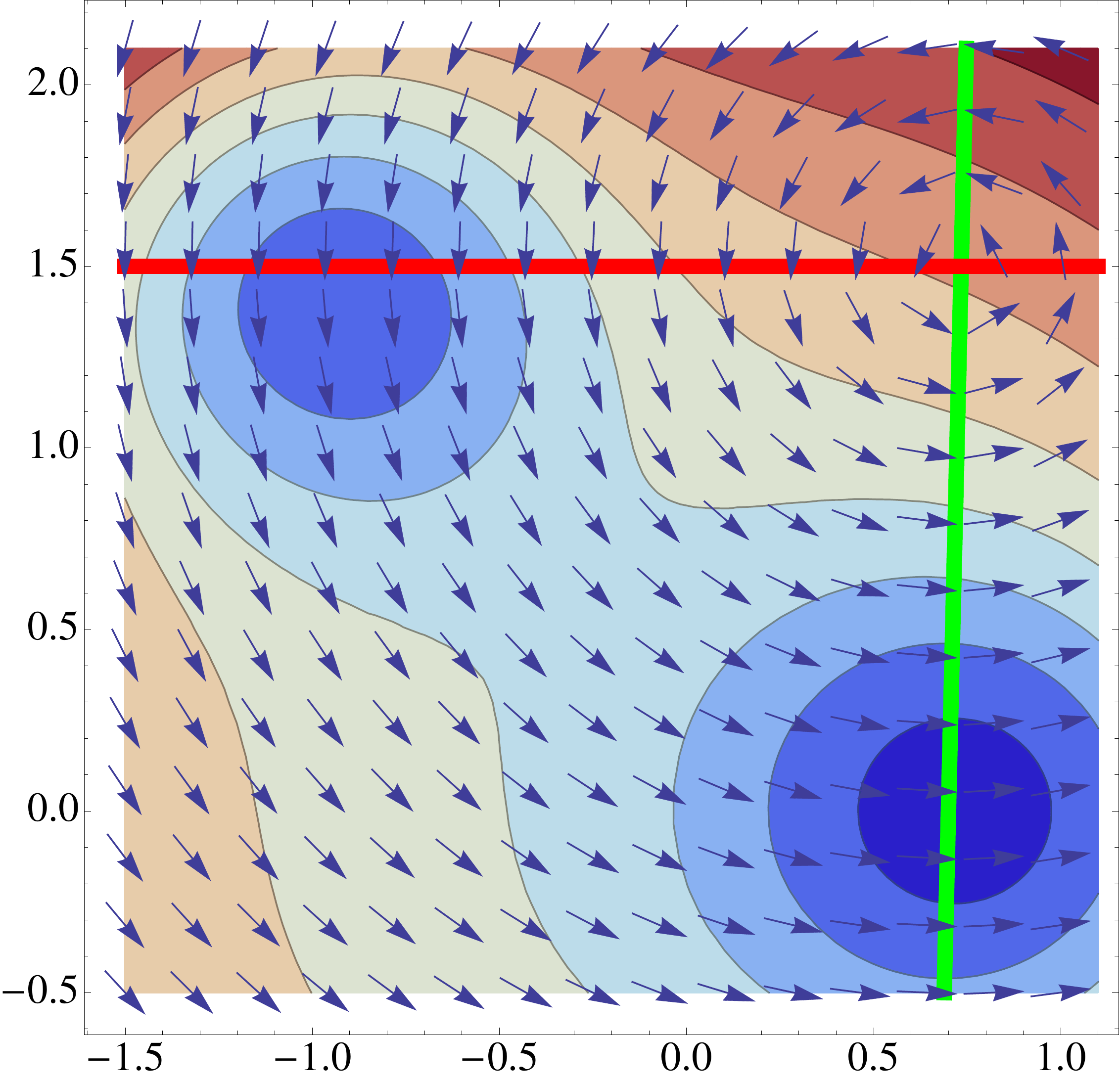,width=3.0in}}
  \caption{Shown here is a representation of the vector field approach to applying $\mathscr{F}_{wind}$ to push milestoning trajectories between two nearly orthogonal planes, subject to our Gaussian potential. The green milestone is defined as the plane for which $\frac{y}{44} - x = -.7$ and the red milestone is defined as the plane for which $y = 1.5$. The vector wind is configured to show the $\mathscr{F}_{wind}$ scheme for accelerating trajectories going from red to green.}
  \label{st}
\end{figure}

\begin{figure}[h]
  \centerline{\epsfig{figure=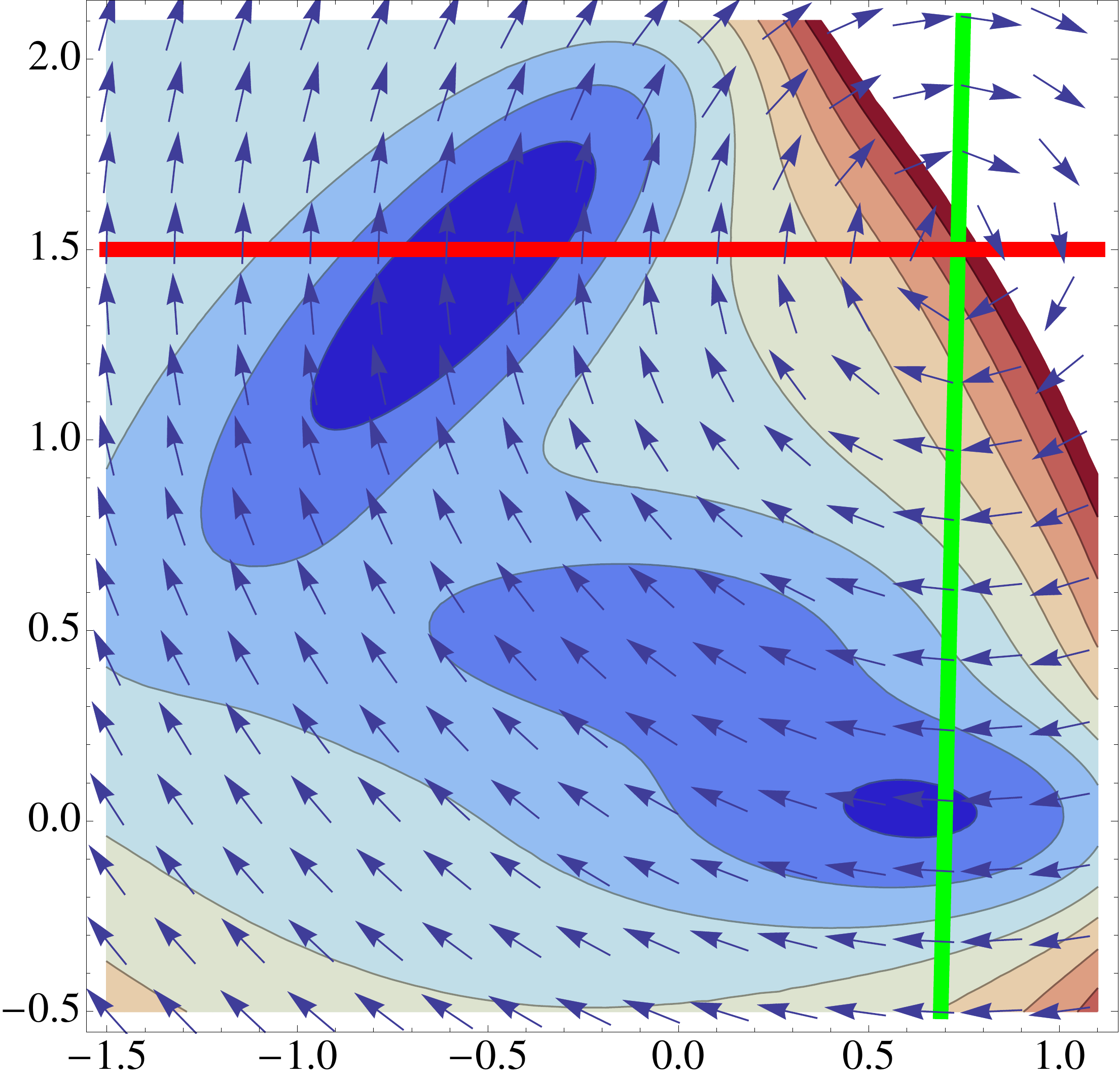,width=3.0in}}
  \caption{This plot shows the same milestone placement and $\mathscr{F}_{wind}$ scheme as the Gaussian potential example applied to the Muller potential and with a directionality for accelerating trajectories from the green milestone to the red one.}
  \label{mt}
\end{figure}

\begin{figure}[h]
  \centerline{\epsfig{figure=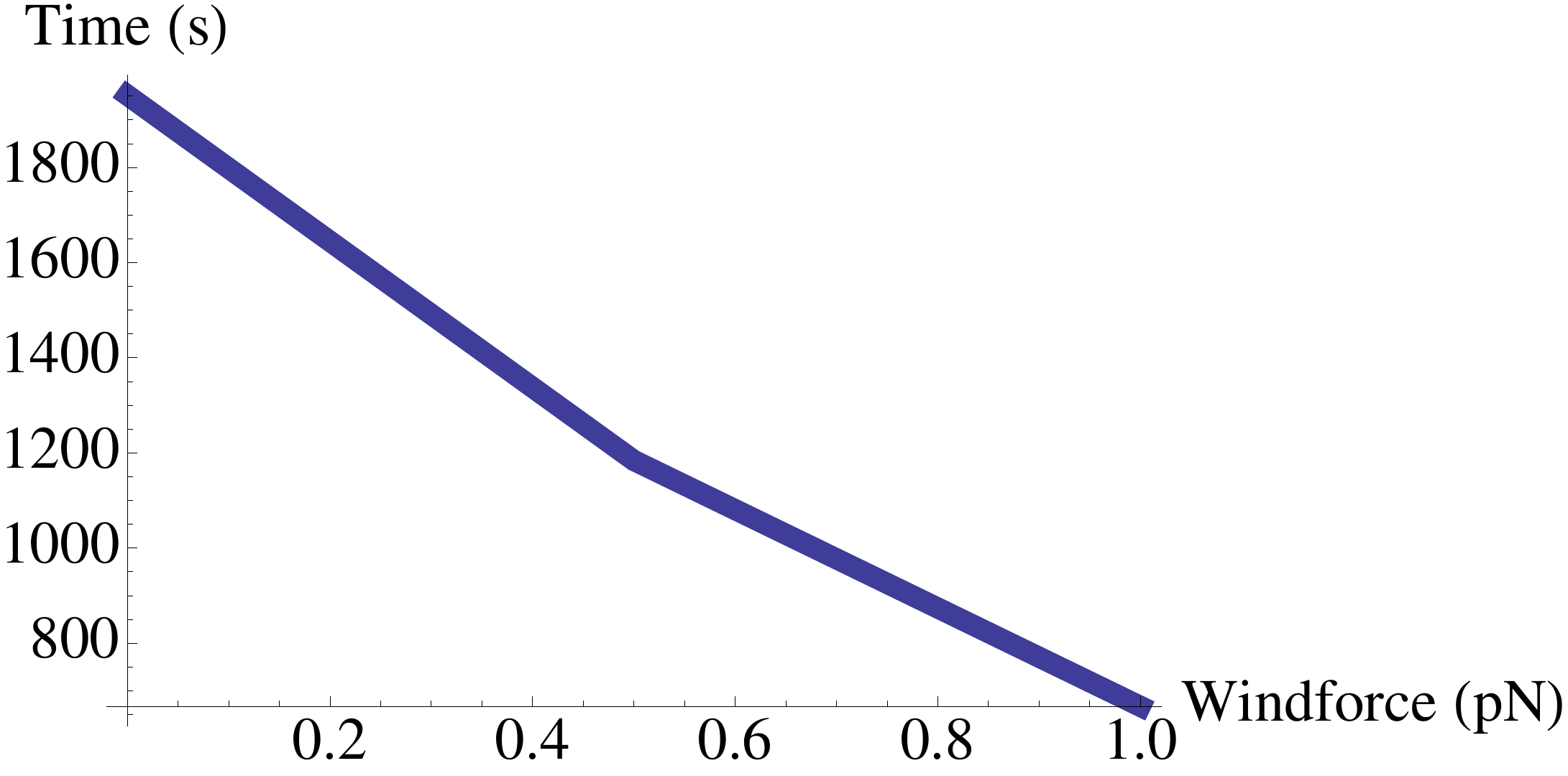,width=3.5in}}
  \caption{This plot shows CPU time as a function of $\mathscr{F}_{wind}$ magnitude for the Gaussian potential.}
  \label{ss}
\end{figure}

\begin{figure}[h]
  \centerline{\epsfig{figure=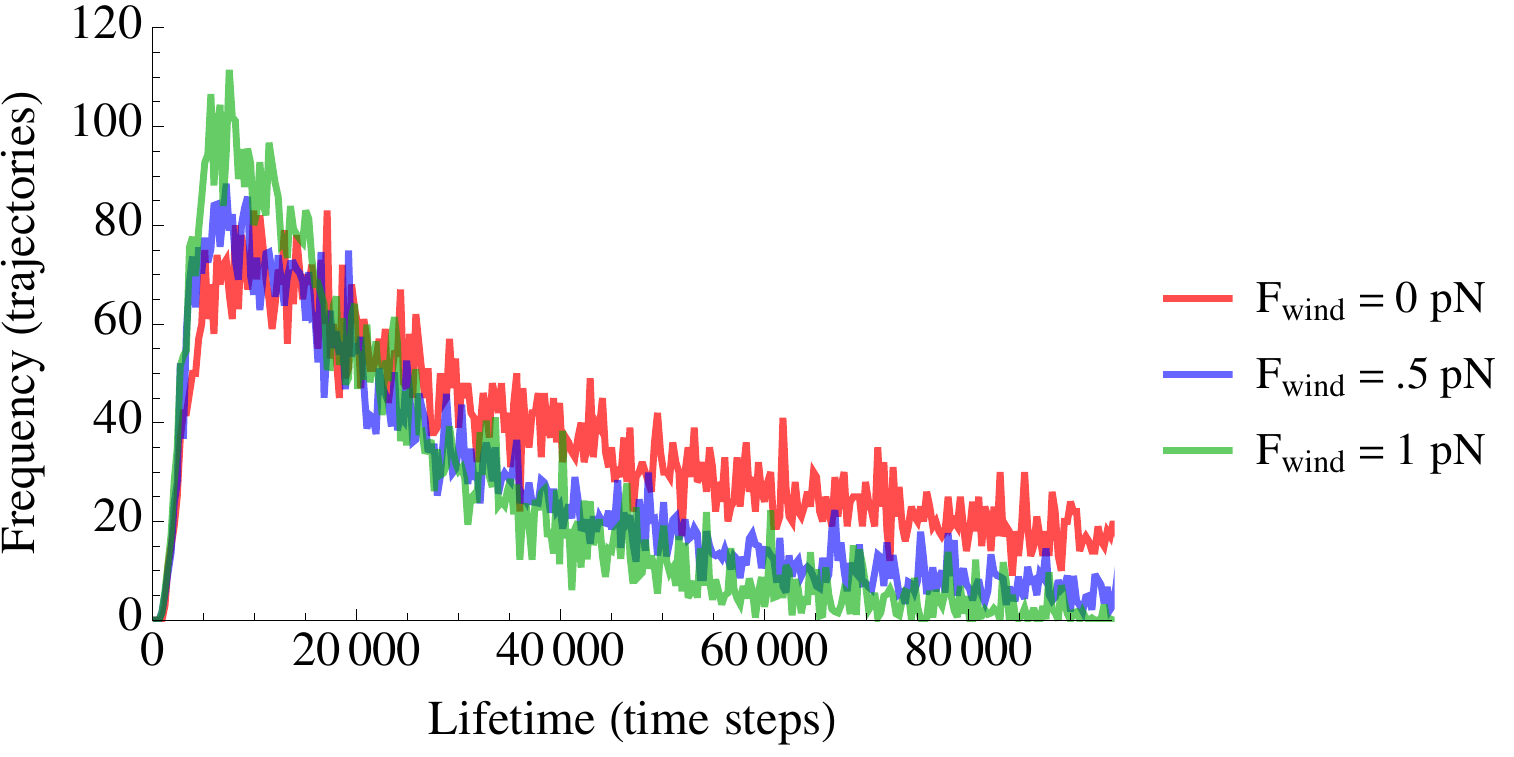,width=3.5in}}
  \caption{This plot shows the $K(\tau)$ functions corresponding to different magnitudes of  $\mathscr{F}_{wind}$ as applied to the Gaussian potential.}
  \label{sk}
\end{figure}

\begin{figure}[h]
  \centerline{\epsfig{figure=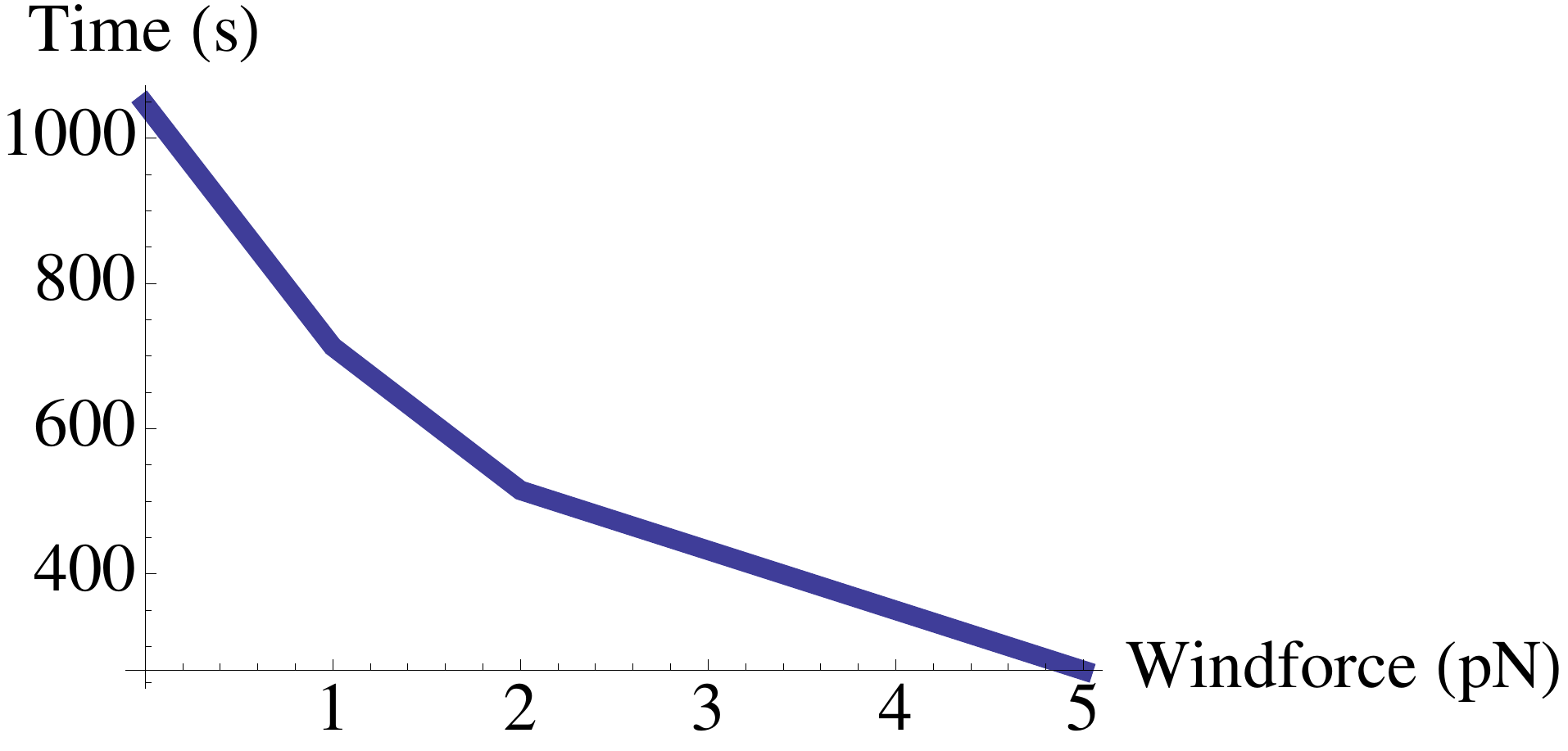,width=3.5in}}
  \caption{This plot shows CPU time as a function of $\mathscr{F}_{wind}$ magnitude for the Muller potential.}
  \label{ms}
\end{figure}

\begin{figure}[h]
  \centerline{\epsfig{figure=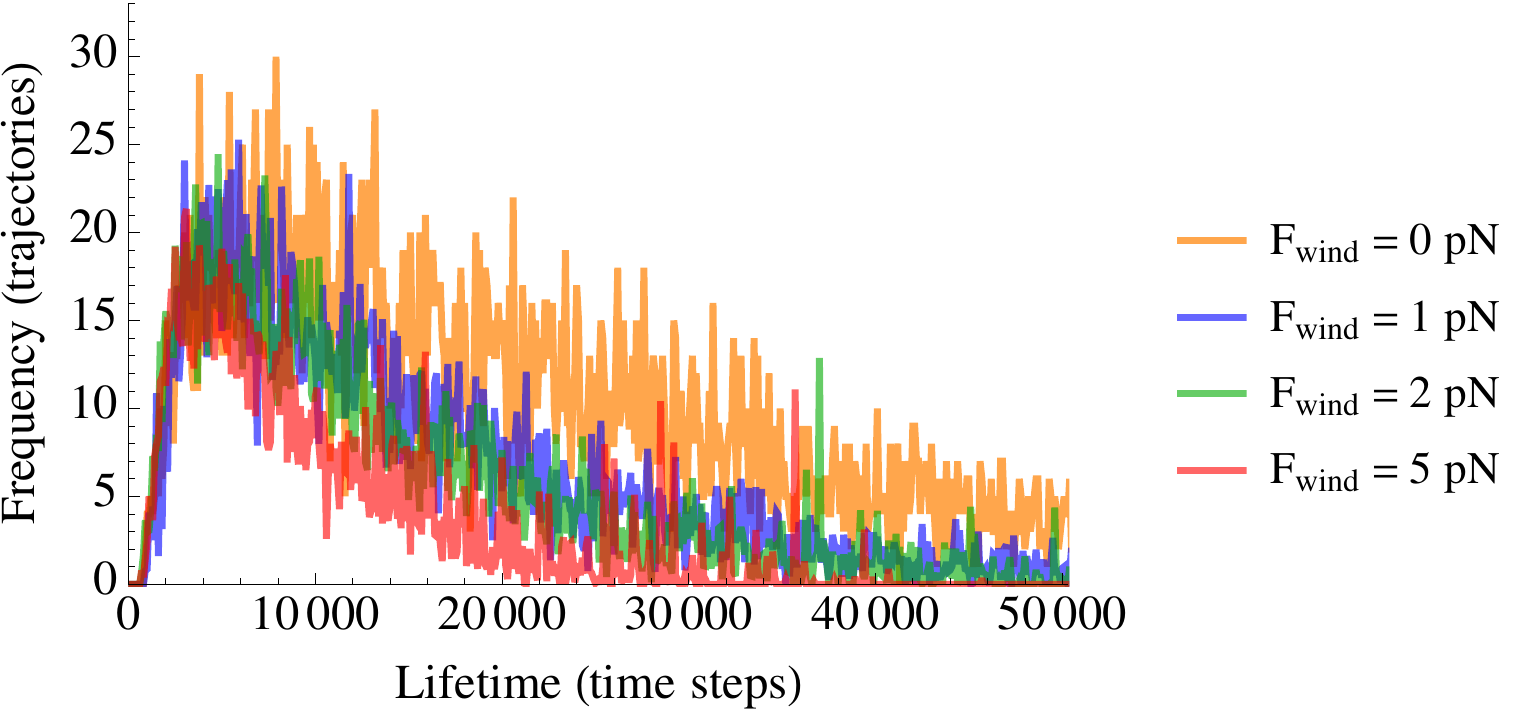,width=3.5in}}
  \caption{This plot shows the $K(\tau)$ functions corresponding to different magnitudes of  $\mathscr{F}_{wind}$ as applied to the Muller potential.}
  \label{mk}
\end{figure}

In all of the preceding examples, $\mathscr{F}_{wind}$ was applied to the system as a constant force applied in a straight line, perpendicular to the parallel milestone hyperplanes, but $\mathscr{F}_{wind}$ can be defined any way we choose. This section demonstrates a method whereby the directionality of $\mathscr{F}_{wind}$ is defined by a vector field which allows $\mathscr{F}_{wind}$ to blow in a curved path between two nearly orthogonal milestones (see figures \ref{st} and \ref{mt}), i.e. our wind has become a tornado! In order to define this vector field, the point of intersection between the two planes was determined, then a function was created which finds the straight line connecting the current position to this point of intersection, and then defines $\mathscr{F}_{wind}$ at that point to be a vector both orthogonal to that straight line and pointing in a clockwise direction. When first passage times were calculated going from the milestone shown in red toward the milestone shown in green (figure \ref{st}), the vector field is simply multiplied by $-1$ to cause our tornado to spin counterclockwise. Using a wind force defined in this manner, we obtain an efficient directionality for $\mathscr{F}_{wind}$ which biases the system toward both leaving its initial milestone in the right direction and approaching its destination milestone, regardless of the positioning of the milestones in configuration space. Another advantage of this scheme is that our curved vector field of $\mathscr{F}_{wind}$ can be defined without any knowledge of the system itself, we only need to know the positions of the milestones, which are always known in milestoning calculations. Figures \ref{st} through \ref{mk} illustrate the application and results of this method using two different 2D potentials, the Muller-Brown potential \cite{muller}, and a simpler Muller-inspired potential with two Gaussian wells we'll call our Gaussian potential. The Gaussian potential is defined as:       

\begin{gather}
V(x, y) = -exp[-(2(x-.8)^2+y^2)] \\ \nonumber
-1.3exp[-((x+1)^2+(y-1.5)^2)] \\ \nonumber
+ .2x^2+.2y^2
\label{gp}
\end{gather}

\noindent and the Muller potential is defined as:

 \begin{gather}
V(x,y) = h \sum_{k = 1}^4 exp[a_k(x-x_k^0)^2 \\ \nonumber
+b_k(x-x_k^0)(y-y_k^0)+c_k(y-y_k^0)^2] \\ \nonumber
\label{mp}
\end{gather}
\noindent where:
\begin{gather*}
A = (-200, -100, -170, 15), a =(-1, -1, -6.5, .7) \\ \nonumber
b = (0, 0, 11, .6), c =(-10, -10, -6.5, .7) \\ \nonumber
x^0 = (1, 0, -.5, -1), y^0 =(0, .5, 1.5, 1) \\ \nonumber
h = .005
\end{gather*}

\noindent A speedup factor of 4 was observed in both the Muller potential and the Gaussian potential, although the $K(\tau)$ functions in the Gaussian potential example displayed less distortion than those produced in the calculations performed using the Muller potential.

\section{Concluding Discussion}
We have presented and tested a method for accelerating milestoning calculations, whereby the true probability density functions for first passage transition time between milestones, $K_{AB}(\tau)$, are recovered from artificially accelerated trajectories via the re-weighting method described in the Theory section. These $K_{AB}(\tau)$ functions are central to milestoning calculations, thus the WARM method presented herein shows potential for broad application.

Our method has been shown to be effective on one and two dimensional potentials with both energetic and entropic barriers, as well as an $11$ dimensional hyperspace, implying that the method should have no scaling limitations, thus the next step will be to test the method on chemical systems. The simplest application would be to apply a single force vector to a single atom which pushes the system toward a configurational change of interest. In this case, the re-weighting factors $S[x(t)]$ and $S_f[x(t)]$ could be calculated by summing the force components in the $x$, $y$, and $z$ directions both with and without the components of the applied force, respectively. 

The main limitation of the WARM method, regardless of the number dimensions are present, is obtaining good re-weighting in the longer $\tau$ range. This is simply a matter of under-sampling. If $\mathscr{F}_{wind}$ is pushing the system to the next milestone so quickly that longer values of $\tau$, relevant to the true $K_{AB}(\tau)$ distribution, are not being sampled, then there just isn't enough density present (or even none at all) to re-weight. This is why the accuracy in the low tau regime is often still quite good when too high of an $\mathscr{F}_{wind}$ has caused the latter portion of the distribution to turn to noise. Thus far, the limitations on the WARM method appear to be solely dependent on whether or not we've pushed the force so hard that trajectories in the longer $\tau$ region of the true $K_{AB}(\tau)$ are even being sampled. For this reason, systems whose true $K_{AB}(\tau)$ distribution functions possess fat tails place the greatest limitations on the degree of computational speedup achievable by WARM. This issue can be addressed in a couple different ways. One approach is to simply define more milestones in the space, the other is to combine the WARM method with some sort of artificial heating method, both modifications which will yield $K_{AB}(\tau)$ functions which decay more rapidly after their peak.

It should be noted that our application of the WARM method to both the high dimensional model, and our vector field-based $\mathscr{F}_{wind}$ implementation of demonstrate that this technique can be applied to systems too complex to intuit the placement of the artificial forces. Given an initial and a final milestone configuration, one could determine the vectors pointing from each atom's initial position to it's final position. Artificial forces, $\mathscr{F}_{wind}$, could then be placed upon all atoms in the system pointing in the direction of these vectors and with a magnitude proportional to the length of the vectors. A zero cutoff could also be added so as not to waste computational resources on applying and accounting for $\mathscr{F}_{wind}$ forces atoms which are beginning at a position fairly close to their destination. We believe that, upon implementation into a molecular dynamics package such as MOIL \cite{Elber1995159}, the WARM method has the potential to be a useful tool for the determination of the kinetic properties of macromolecules.

\section{Acknowledgments}  IA acknowledges funds from an NSF CAREER award (CHE-0548047).

\bibliographystyle{unsrt}
\bibliography{WARM}

\end{document}